\documentclass[11pt,a4paper]{article}
\usepackage{jcappub}
\usepackage{amssymb,amsmath,amsfonts,palatino,amsthm}

\usepackage{slashed}
\usepackage{amsfonts}
\usepackage{amsmath}
\usepackage{amssymb}
\usepackage{bm}
\usepackage{dcolumn}
\usepackage{epsfig}
\usepackage{graphicx}
\usepackage{graphics}
\usepackage[latin1]{inputenc}
\usepackage{latexsym}
\usepackage{rotating}
\usepackage{hyperref}
\usepackage{floatrow}

\newcommand\be{\begin{equation}}
\newcommand\ba{\begin{eqnarray}}
\newcommand\ee{\end{equation}}
\newcommand\ea{\end{eqnarray}}

\title{Dynamics of Gauge Field Inflation}

\author[a]{Stephon Alexander}
\author[a]{Dhrubo Jyoti}
\author[b,c]{Arthur Kosowsky}
\author[d]{Antonino Marcian\`o}

\affiliation[a]{Center for Cosmic Origins and Department of Physics and Astronomy, 6127 Wilder Laboratory, Dartmouth College, Hanover, NH 03755, USA}
\affiliation[b]{Department of Physics and Astronomy, University of Pittsburgh, Pittsburgh, PA 15260, USA}
\affiliation[c]{Pittsburgh Particle Physics, Astrophysics, and Cosmology Center (Pitt-PACC), Pittsburgh, PA 15260, USA}
\affiliation[d]{Center for Field Theory and Particle Physics \& Department of Physics, Fudan University, 200433 Shanghai, China}

\emailAdd{dhrubo.jyoti@dartmouth.edu}
\emailAdd{stephon.alexander@dartmouth.edu}
\emailAdd{kosowsky@pitt.edu}
\emailAdd{marciano@fudan.edu.cn}
\arxivnumber{1408.4118}

\abstract{
We analyze the existence and stability of dynamical attractor solutions for cosmological inflation driven by 
the coupling between fermions and a gauge field.  Assuming a spatially homogeneous and isotropic gauge field and fermion current,  
the interacting fermion equation of motion reduces to that of a free fermion up to a phase shift. 
Consistency of the model is ensured via the St\"uckelberg mechanism.
We prove the existence of exactly one stable solution, and demonstrate the stability numerically. Inflation arises without
fine tuning, and does not require postulating any effective potential or non-standard coupling.}

\begin{document}

\maketitle

\section{Introduction}
\noindent Inflation in the early universe stands as a cornerstone of the standard model of cosmology, providing a simple physical mechanism for understanding several fundamental properties of the universe which otherwise are puzzling. Along with naturally explaining the observed flatness and isotropy of the universe, and the lack of monopoles from breaking a higher gauge symmetry to the standard model gauge group, inflation provides a profound quantum mechanical origin for small density perturbations over a wide range of super-horizon scales which have Gaussian random statistics and equal fractional amplitude in all particle species. Recent measurements of the cosmic microwave background temperature power spectrum \cite{WMAP9,ACT,SPT,Planck} show that, to an excellent approximation, this is the universe we live in. 

The fundamental physical basis for inflation, however, remains obscure. Inflation was first developed by borrowing concepts from condensed matter and particle physics. In the original inflation proposal by Guth \cite{Guth:1980zm}, both parity-symmetry breaking and SU$(5)$-symmetric unified theories were essential ingredients. Soon, because of the amount of fine tuning required by the earliest model of inflation (so-called ``Old Inflation'') and difficulty making inflation end, the focus of inflation theories shifted from the original goal of making contact with fundamental physics concepts to the consideration of effective models described by new scalar fields. Although phenomenologically successful, this paradigm has a high degree of arbitrariness in the choice of the potential for the inflaton scalar field, along with other difficult issues \cite{Brandenberger:2008zz}. A notable model of inflation based on the only known fundamental scalar field, the Higgs field, has been constructed \cite{Bezrukov:2007ep}, but at the cost of introducing a non-standard coupling to the gravitational field \cite{GerKe}. 

Conceptual alternatives to scalar-field inflation have also been considered, such as inflation driven by vector fields \cite{Ford:1989me,Golovnev:2008cf,Maleknejad:2011sq}. (For a recent review on gauge fields and inflation, see \cite{Jabbari0}.) In this paper we consider another idea proposed by Alexander, Marcian\`o and Spergel \cite{CSI}, based on the interaction of an abelian gauge field with a fermion current. In this case, inflation is driven by the energy density of interacting fermion and gauge fields, rather than the potential energy of a scalar field, and invokes only the standard interaction Lagrangian between gauge fields and a fermion current. Our specific model differs from that in Ref.~\cite{CSI}: we use the St\"uckelberg mechanism to provide a self-consistent model with gauge invariance. It is similar in spirit to the inflation model of Ref.~\cite{Adshead:2012kp}, which uses interacting gauge and axion fields.

Here we assume that the early universe was dominated by a time-like homogeneous gauge field and a fermion field. Classical physics intuition suggests that, in an expanding universe, both   fermion charge density (the total charge density of fermions minus the total charge density of anti-fermions) and electromagnetic fields should dilute. However, the coincident-point limit of the fermion propagator in a quasi-deSitter space-time with constant deceleration is actually constant instead of decaying with redshift \cite{Koksma:2009tc}, and we may show that the vacuum expectation value (VEV) of the fermion current is as well, namely $J^\mu\propto\,\delta^\mu_0$ \cite{CSI}. Using this result to solve consistently for the gauge field, we show that the gauge field $A^t$ also remains constant. In other words, due to quantum effects of the expanding space-time, the time-like component of the gauge field does not dilute. This implies that the gauge field and fermion current interaction energy $A\!\cdot\!\mathcal{J}$ is nearly constant. This interaction energy can therefore sustain an inflationary phase of the universe. 

Building on the work done in \cite{CSI}, we investigate the existence of attractor solutions in this theory. We begin by introducing the action in Sec.~II, which invokes the St\"uckelberg mechanism to provide a self-consistent model. We derive the equations of motion of the gauge (Sec.~III) and the fermion fields (Sec.~IV) in the background expanding space-time with the usual assumptions of isotropy and spatial homogeneity. We prove the conservation of energy-momentum in Sec.~V. In Sec.~VI, we incorporate the Friedmann equation for the space-time expansion dynamics, giving a 3-dimensional dynamical system, and find a fixed point corresponding to inflationary behavior. In Sec.~VII, we perform a stability analysis by expanding the system to first order in the perturbations about the fixed point, and present a numerical realization of our model with a phase portrait. The inflationary solution has a broad basin of attraction, demonstrating that inflation is generic in this model. We then conclude with some thoughts about how this model relates to more conventional inflation theories, how perturbations may contribute to ending inflation, and open questions about the connection between inflation and the microphysical interactions driving it. 

\section{The Theoretical Model}

We consider a U$(1)$ gauge field $A_\mu$ with mass $m$ which interacts with a Dirac fermion $\psi$ of mass $M$. We use the St\"uckelberg mechanism to allow for gauge invariance of $A_\mu$. Integration is over a four-dimensional space-time manifold $\mathcal{M}_4$ with Lorentzian signature. The metric field is denoted as $g_{\mu \nu}$, with Greek indices labeling space-time coordinates, and its determinant as $g$. The Ricci scalar is denoted as $R$ and the Planck scale as $M_p^2=G^{-1}$ in natural units. In the action for the Dirac field, we use the conventional notation $\slashed{\nabla}$ for the contraction of the covariant derivative (with respect to the spin-connection) $\nabla_\mu$ with the Dirac matrices $\gamma^I$, through the inverse tetrad field $e_I^\mu$. Finally, $I$ stands for an internal index from $1$ to $4$. The action is then written as  
\be\label{Action}
S=\int_{\mathcal{M}_4} d^4x \sqrt{|g|} \left[\frac{M_p^2R}{8\pi}-\frac{1}{4}G_{\alpha\beta}G^{\alpha\beta}-\frac{1}{2}m^{2}C_{\mu}C^{\mu}+C_\mu\mathcal{J}^\mu+\mathcal{L}_{\rm fer}\right]\,,
\ee
\small
\be
C_\mu=A_\mu-\frac{1}{m}\partial_\mu\theta\,,~~~G_{\mu\nu}=\partial_{[\mu}C_{\nu]}=\partial_{[\mu}A_{\nu]}=F_{\mu\nu}\,,~~~\mathcal{L}_{\rm fer}=-\imath \overline{\psi}\slashed{\nabla}\psi + c.c. + M\overline{\psi}\psi\,,~~~\mathcal{J}^\mu=q\,\bar{\psi}\gamma^\mu\psi\,,\nonumber \ee
\normalsize
where $q$ is a dimensionless coupling constant. Gauge invariance is ensured by the St\"uckelberg scalar field $\theta$, and by the transformations 
\ba
&&A_\mu\rightarrow A_\mu'=A_\mu\,+\,\,\partial_\mu\Lambda\,,\nonumber\\
&&~\theta~~\rightarrow ~\,\theta'\,=~\,\theta~\,+\,\,m\,\Lambda\,.
\ea
A generalization of this theory to  non-abelian SU(N) Yang-Mills fields will be considered elsewhere \cite{AMM2}.

\section{Equation of Motion of the U$(1)$ Field}

The sector of the action involving the U$(1)$  gauge field is given by
\be
\mathcal{L}_{U(1)}=\sqrt{-g} \left[-\frac{1}{4}F_{\alpha\beta}F_{\alpha'\beta'}
\,g^{\alpha\alpha'}g^{\beta\beta'}-\frac{1}{2}m^{2}\left(A_\mu-\frac{1}{m}\partial_\mu\theta\right)^2\,+J^I e_I^\mu A_{\mu} \right]\,,
\ee
where
\be\label{JI}
J^I=\bar{\psi}\,\gamma^I \psi \,,
\ee
and in which for convenience we set the dimensionless coupling constant $q$ to $1$. The equations of motion are straightforward to obtain:
\be\label{eom0}
\partial_\mu\left[\sqrt{-g}\left(-F^{\mu\nu}\right)\right]=\sqrt{-g}\left(-m^2A^\nu+m\,\partial^\nu\theta+\mathcal{J}^\nu\right)\,.
\ee
We impose the generalized gauge condition \cite[Eq.~16]{Ruegg}\,,
\be\label{gauge}
\nabla_\mu A^\mu=m\theta\,,
\ee
which relates the St\"uckelberg scalar field $\theta$ to the longitudinal mode of the massive U$(1)$ gauge field $A_\mu$. In the standard Friedmann-Lema\^itre-Robertson-Walker (FLRW) metric  $ds^2\!=-dt^2+a^2(t)d{\bf x}^{\,2}$ and the time derivative denoted with a ``$\dot{\phantom{a}}$'', the vacuum expectation value (VEV) of the fermion current on a quasi-de Sitter background has the form \footnote{The propagator in \cite{Koksma:2009tc} was derived assuming that $M/H$ and $\epsilon$ are constant. The former condition can be satisfied by giving mass to the fermion via a Yukawa coupling, and the latter condition becomes unimportant in the phase space near to the attractor solution. In other words, the result \eqref{Current} was shown in \cite{CSI} to be strictly true only in the limit of small $\epsilon$. For simplicity, we will address these subtleties elsewhere, and instead focus only on the basic dynamics.}
\be\label{Current}
\mathcal{J}^\sigma\simeq \frac{J\,\delta_0^{\sigma}}{1-\epsilon}\,, ~~~~\epsilon\neq1\,,
\ee
where $\epsilon$ is the deceleration parameter defined as
\be\label{dec}
\epsilon\equiv -\frac{\dot{H}}{H^2}\,,
\ee 
and the constant amplitude $J$ was derived in \cite[Eqs.~5.5]{CSI}. Indeed, the expectation value of $\mathcal{J}^\sigma$ in the Bunch-Davies vacuum is constant for the temporal component and vanishes for the spatial components. 
This can be easily checked at $\epsilon=0$ from the fermion propagator $S^{ab}(x, y)$, which reads
\ba \label{fp}
&&\imath S^{ab} (x, y) = a(\eta_x) ( \imath \gamma^\mu \nabla_\mu + M) \frac{H^2}{\sqrt{a(\eta_x) a(\eta_y) }}\, \left[ \imath S_+(x,\,y) \frac{1+\gamma^0}{2} + \imath S_-(x,\,y) \frac{1-\gamma^0}{2}       \right]   \,, \nonumber\\
&& \imath S_\pm(x,\,y) = \frac{\Gamma(1\mp \imath \frac{M}{H} ) \, \Gamma(2 \pm \imath \frac{M}{H} )}{(4\pi)^2 \Gamma(2)}\!\! \phantom{a}_2 F_1\!\left( 1 \mp \imath \frac{M}{H}, \, 2\pm \imath  \frac{M}{H}, \, 2, 1-  \frac{\Delta(x,y)}{4} \right)\,, \nonumber\\
&& \Delta(x,y)= a(x) a(y) H^2 \Delta x^2\,, \qquad \Delta x^2= (|\eta_x -\eta_y| - \imath \varepsilon)^2 + ({\bf x} - {\bf y}) \cdot ({\bf x} - {\bf y})\,. 
\ea
The coincidence limit of the latter expression, expanded to first order in $M/H$, is then
\be \label{ja}
\lim_{y \rightarrow x} S^{ab}(x,y)\simeq \frac{H^2}{16 \pi^2} \left[ \frac{M^2}{H} \, \left( \frac{1+\gamma^0}{2} \right)^{ab} - \frac{M^2}{H} \,  \left( \frac{1+\gamma^0}{2} \right)^{ab} \right] + \mathcal{O}\left( \frac{M^2}{H^2} \right)\,.
\ee
The comoving fermion number density, which is obtained by multiplication of the physical fermion number density 
\be
\langle 0|  \mathcal{J}^0 |0\rangle\simeq \lim_{y \rightarrow x} S^{ab}(x,y) \, \gamma^0_{ab}
\ee 
times $a^3$, increases as the universe expands and thus amounts to the creation of particles. The unfamiliar time evolution in (\ref{ja}) and in (\ref{Current}) is the rigorous result of the expectation value of second-quantized field bilinears in the Bunch-Davies vacuum. On the other hand, from the Dirac equation it is straightforward to show the
usual semiclassical result $\mathcal{J}^0=\overline{\psi} \gamma^0 \psi \propto 1/a^3$. 
So the creation of particles in a quasi-de Sitter spacetime is purely a property of quantized particles in curved spacetime. 

In Equation \eqref{Current} $J$  is a somewhat complicated function of $M$ and $\bar H$ \cite[Eqs.~4.18]{CSI}, where $\bar H$ is the de Sitter limit of the model, assuming it exists. Obtaining $J$ requires solving the Friedmann equation. Regardless of the specific value of $J$, it is constant in time which is necessary to have an inflationary solution with constant energy density.

Note that the divergence of the fermion current does not vanish, $\nabla_\mu\mathcal{J}^\mu\neq0$. As a result, taking the divergence of both sides of the equation of motion \eqref{eom0} shows that the massless limit $m\rightarrow 0$ does not exist. A massless gauge field will be inconsistent with the assumed form for the current; this is the reason the gauge field includes a mass term in the Lagrangian \eqref{Action}.    

As a starting point for an inflation model, assume that $A^\mu$ is spatially homogeneous, \emph{i.e.} $A_\mu=A_\mu(t)$. From \eqref{eom0}, we obtain the equation of motion for the space-like components
\be\label{space}
\ddot{A_i}+m^2A_i=0.
\ee
Solving for $A^i$, we can easily see that the energy density contribution from the space-like gauge fields, $m^2A_{i}A^{i}$,  dilutes as $1/a^2$ like a classical magnetic field. We may therefore set $A_i=0$, and they do not play any further role. For the time-like component,
\be\label{eom}
\ddot{A_0}+3H\dot{A_0}+3\dot{H}A_{0}+m^{2}A_0+J/(1-\epsilon)=0\,.
\ee
To solve for $A_0$ also requires the Friedmann equation for $H(t)$. In the limit of ideal de Sitter phase ($H(t)={\bar H},$ $\epsilon=0$) the equation of motion \eqref{eom} reduces to that of a damped Harmonic oscillator with a constant driving term $J$. The steady-state solution is
\be\label{sol}
A^0=\frac{J}{m^{2}}\,,
\ee
which gives the constant energy inflationary solution $\rho\simeq\frac{1}{2}m^2 A^2+ A \cdot\mathcal{J}+MJ$. We will show that this limit is a dynamical attractor
in Sec.~VII. 

The behavior of the St\"uckelberg scalar field immediately follows from the gauge condition \eqref{gauge},
\be\label{theta}
\theta=\frac{1}{m}\nabla_\mu A^\mu=\frac{1}{m}\nabla_0\, A^0=\frac{1}{m}\left(\dot{A^{0}}+3{\bar H}A^0\right)=\frac{3{\bar H}J}{m^3}\,.
\ee
That is, the scalar field is constant in the steady-state limit. Therefore, since only derivatives of the scalar field $\partial_\mu\theta$ appear in the Lagrangian \eqref{Action}, the scalar field does not contribute any energy density or pressure in the steady-state limit. The solution \eqref{scalar} is also consistent with the scalar field equation of motion,
\be\label{scalar}
\ddot{\theta}+3H\dot{\theta}+m^2\theta=\frac{J}{m\left(1-\epsilon\right)}\left[3H+\frac{\dot{\epsilon}}{\left(1-\epsilon\right)}\,\right].
\ee
This completes our calculation of the equations of motion of our gauge and scalar fields. 

\section{Fermion Equation of Motion: Free Versus Interacting}\label{FermionEoM}

The fermion current vacuum expectation value \eqref{Current} was derived in \cite{CSI} using the \emph{free} fermion propagator worked out in \cite{Koksma:2009tc}. 
Here we reduce the full interacting propagator to the free propagator. 
The equation of motion for the four-spinor $\psi\left(x^\mu\right)$ results from varying the action \eqref{Action} with respect to $\overline{\psi}\equiv\psi^\dagger\gamma^0$\,,
\be\label{eom2}
\imath \gamma^\mu\nabla_\mu\psi- m\psi - q\gamma^I e_I^\mu \psi C_\mu=0\,,
\ee
where for the gravitational covariant derivative $\nabla_\mu=\partial_\mu-\Gamma_\mu$. Using the FLRW vierbein and its inverse 
\ba
&& e_\mu^I=a(\eta)\delta_\mu^I \nonumber\,, \\
&& e_I^\mu=a^{-1}(\eta)\delta_\mu^I\,,
\ea
where $\eta$ is the conformal time for which FLRW metrics read $ds^2=a^2(\eta)(d\eta^2-d\vec{x}^{\,2})$, and the definition of the spin-connection
\be
\Gamma_\mu=-\frac{1}{8}e_I^\nu\left(\partial_\mu e_{\nu J}-\Gamma_{\mu\nu}^\alpha e_{\alpha J}\right)\left[\gamma^I,\gamma^J\right]\,,
\ee
it can be shown that 
\be
\imath \gamma^\mu\nabla_\mu\psi=\imath a^{-5/2}\gamma^I\partial_I\left(a^{3/2}\psi\right)\,,
\ee
where the covariant derivatives have been conveniently replaced with partial derivatives in the tangent space. We use capitalized Latin letters as opposed to Greek letters to indicate the tangent space. The tangent space is simply Minkowski space, in which $\gamma^I$ are constant matrices. The above relation gives
\be
- q\gamma^I e_I^\mu \psi C_\mu=-qa^{-1}\gamma^0\psi C_0\,,
\ee
which again assumes that only the time-like gauge field is non-vanishing. Now define the rescaled four-spinor as
\be\label{densitize}
\chi=a^{3/2}\psi\,,
\ee
and recast equation \eqref{eom2} as
\be\label{eom3}
\imath \gamma^I\partial_I\chi-ma\chi-q\gamma^0\chi C_0=0\,.
\ee
Equation \eqref{eom3} can be written as
\be\label{withA0}
-\imath \gamma^0\partial_\eta \chi+\imath \gamma^i\partial_i\chi-ma\chi-q\gamma^0\chi C_0=0.
\ee
This is the full interacting fermion equation of motion for the model we are considering. However,  the interaction term $-q\gamma^0\chi C_0$ causes only a global phase shift in the spinor. Assuming that $C_0$ is a well-behaved function, some function $F(\eta)$ will exist such that 
\be
\partial_\eta F(\eta)\equiv C_0\,.
\ee
That is, $F$ is the anti-derivative of $C_0$. Consider
\be\label{shift}
\tilde{\chi}=\chi\,e^{\imath qF(\eta)}\,;
\ee
substituting \eqref{shift} into \eqref{withA0}, we have
\be\label{freeferm}
-\imath \gamma^0\partial_\eta \tilde{\chi}+\imath \gamma^i\partial_i\tilde{\chi}-ma\tilde{\chi}=0.
\ee
This is sufficient to show that the presence of the interaction term $q\gamma^I e_I^\mu \psi C_\mu$ in the equation of motion (\ref{withA0}) is equivalent to a phase shift of the free fermion. Since the fermion propagator has the form $\bar{\psi}\,\psi$, this phase shift has no effect on the propagator. Thus, the interacting fermion propagator must be identical to the free fermion propagator derived in \cite{Koksma:2009tc}.  This ensures that the use of the result \eqref{Current}, derived in \cite{CSI} based on the free fermion propagator recovered in \cite{Koksma:2009tc}, is justified.

\section{Conservation of Energy-Momentum}\label{dyn}

The energy-momentum tensor for the theory specified in Eq.~(\ref{Action})  reads
\begin{eqnarray} \label{tensor}
T^{\mu\nu}
= G_\alpha^{\,\,\mu} G^{\alpha\nu} 
-\frac{1}{4} g^{\mu\nu}  G_{\alpha \beta} G^{\alpha \beta}-\frac{1}{2}m^2g^{\mu\nu}C_\alpha C^\alpha-\,g^{\mu\nu} C_\rho \mathcal{J}^\rho +C^{(\nu} \mathcal{J}^{\mu)}\,+\,^{(\psi)}T^{\mu\nu},
\end{eqnarray}
where $\,^{(\psi)}T^{\mu\nu}$ denotes the fermion contribution,
\be
^{(\psi)}T^{\mu\nu}=\frac{\imath}{2}\left[\bar{\psi}\,\Gamma^{\,(\mu}\,\nabla^{\,\nu)}\psi-\nabla^{(\nu}\,\bar{\psi}\,\Gamma^{\mu)}\psi\right]-\,g^{\mu\nu}\left[\frac{\imath}{2}\left(\bar{\psi}\,\Gamma^{\mu}\,\nabla_{\mu}\psi-\nabla^{\mu}\,\bar{\psi}\,\Gamma_{\mu}\psi\right)+M\bar{\psi}\psi\right]\,,
\ee
where $\Gamma^\mu=\gamma^I e^\mu_I$. In a FLRW space-time, the energy-momentum tensor has the form $T^{\mu\nu}={\rm diag}(\rho,\,P,\,P,\,P)$. Clearly \eqref{tensor} is diagonal due to the assumed isotropy and homogeneity of the fields. We can show the local conservation of energy-momentum in the steady-state solution \eqref{sol} and \eqref{theta} as follows. Because of isotropy of the current, the spinor must have the form $\psi=(\psi_0,0,0,0)$ in the standard Dirac basis \cite{ArmendarizPicon:2003qk}. Using \eqref{JI}, we  deduce the component of the energy density due to fermions, 
\be
\rho^{(\psi)}=M\bar{\psi}\psi=MJ\,,
\ee
The total energy density is  therefore 
\be
\rho= T_{00}= \frac{1}{2}m^2\mathcal{C}^2+\,\,\mathcal{C}\cdot\mathcal{J}+MJ\,,
\ee
and the total pressure 
\be
P=T_{ii}=-\frac{1}{2}m^2\mathcal{C}^2-\,\,\mathcal{C}\cdot\mathcal{J}-MJ\,.
\ee
From the expressions for the energy density and pressure, it is then straightforward to check the conservation equation
in comoving coordinates,
\be\label{conservation}
\nabla_\mu T^{\mu\nu}=0 \implies\dot{\rho}+3H(\rho+P)=0\,,
\ee
as follows. For a de Sitter background space-time, $C_\mu=\delta_\mu^0\,J/m^2$ follows from \eqref{sol} and \eqref{theta}. Thus, the $C_\mu$ contribution is constant, and 
$\dot{\rho}=0$. Substituting these expressions for $\dot{\rho}$, $\rho$ and $P$ into  \eqref{conservation} shows that the conservation equation is satisfied.

Note that the assumption of spatial isotropy of the gauge fields $A^\mu$ gives a vanishing magnetic field, {\it i.e.} ${\bf B}=0$. In addition,  since $A^i=0$, also  the electric field 
${\bf E}=0$. Essentially, the four-current \eqref{Current} gives rise to a spatially uniform electric potential $A^0$ that permeates the universe and stays constant despite cosmic expansion. This creates a constant energy density $\frac{1}{2}m^2 A^2+ A \cdot\mathcal{J}+M\,J$ that can sustain an inflationary phase of the universe.

\section{Dynamical System}

The total energy density is
\be\label{energydensity}
\rho=\frac{1}{2}m^2(A_\mu-\frac{1}{m}\partial_\mu\theta)^2-\,\,(A_\mu-\frac{1}{m}\partial_\mu\theta)J^{\mu}+MJ\,.
\ee
Then the first Friedmann equation is simply  $H^2=8\pi\rho/\left(3{M_p}^2\right)$, assuming a spatially-flat universe. Within comoving coordinates, substituting in $\rho$ from \eqref{energydensity} gives
\be\label{H}
H^2=\frac{8\pi}{3M_p^2}\left[\frac{1}{2}m^2{A^0}^2+A^0\left\{\frac{ J}{\left(1-\epsilon\right)}-m\dot{\theta}\right\}+\frac{1}{2}\dot{\theta}^2+MJ\right],
\ee
using \eqref{Current}. For the steady-state solution \eqref{sol}, we have
\be\label{H0}
{\bar H}^2=\frac{8\pi}{3M_p^2}\left[\frac{3}{2}\frac{J^2}{m^2}+MJ\right]\,.
\ee
We would like to build a model where the gauge field and its interaction with the fermion field is the dominant source of energy density, as opposed to the energy contribution of the fermions themselves. 
Therefore we choose the two physical parameters $m$ and $M$ such that  $J^2/m^2 \gg MJ$ in \eqref{H0} , so that 
\be\label{gaugefielddom}
\frac{J}{Mm^2}\gg 1\,.
\ee
In this limit, we have,
\be\label{H02}
{\bar H}^2\simeq\frac{4\pi J^2}{M_p^2m^2}\,.
\ee
In \cite{CSI},  $J$ was  computed in the limit $M/{\bar H}\ll1$  to be $J\sim M^2{\bar H}/4\pi^2 +\mathcal{O}(M/{\bar H})^3$. We can easily see in \eqref{H02} that substituting in $J=\frac{1}{4\pi^2}M^2{\bar H}$ gives no unique solution for ${\bar H}$ since it cancels out from both sides of the equation. 
We deduce that the limit $M/{\bar H}\ll1$ is inconsistent with our model. 
Instead, assume that
\be\label{g1}
M/{\bar H}\geq1\,.
\ee
In that case, we expect from \cite[Eqs.~4.18]{CSI} that $J$ is a complicated function of $M$ and ${\bar H}$. Nevertheless, the function $J(M,{\bar H})$ will still be constant in time, allowing an inflationary solution. Given the function $J(M,{\bar H})$, we can plug it into \eqref{H02} and check if a unique solution for ${\bar H}$ exists for some value of $m$, the other independent parameter in our model.
Consider the choice ${\bar H}=M$. Then, by dimensional considerations,  we must have $J\simeq M^3$.  Plugging into \eqref{H02} gives $m\simeq M^2/M_p$. We suspect
that ${\bar H}$ should be GUT-scale, ${\bar H}\simeq 10^{-3} M_p$. Thus,  we have a GUT-scale $M$, and a sub-GUT scale $m$, $m\simeq 10^{-6}M_p$, which are large but 
plausible values. Crucially,  \eqref{gaugefielddom} is satisfied, {\it i.e.} the fermion energy contribution is subdominant. From \eqref{sol} follows $A^0=J/m^2\simeq M_p^2/M=10^3 M_p$\,, {\it  i.e.} the gauge field is super-Planckian. But this is fine physically since the scale of the energy density, which is equal to ${\bar H}$, is sub-Planckian. This model can therefore be described as a large-field model of inflation. 
Solutions of ${\bar H}$ in terms of $M$ and $m$ are non-trivial in the general case \eqref{g1}. Nevertheless, at least one consistent solution exists for $M={\bar H}$.  
We therefore proceed a dynamical analysis assuming that $J$ exists, leaving the explicit calculation of the general case for future work.

It is convenient to use the dimensionless dynamical variables
\be\label{adim}
\tilde{t}\equiv t\,m,\qquad  X\equiv\frac{m^2}{J}\,A^0, \qquad\Theta\equiv\frac{m^2}{J}\,\theta, \qquad \mathcal{H}\equiv \,H/m\,.
\ee
The characteristic time scale is $m^{-1}$. First rearrange the equation \eqref{eom} as 
\be\label{eoms}
{X}''+3\mathcal{H}{X}'+(3\dot{\mathcal{H}}+1)X=1/\left(1+\dot{\mathcal{H}}/\mathcal{H}^2\right)\,,
\ee 
where we use a prime to indicate derivative with respect to $\tilde{t}$.
The dimensionless Friedmann equation follows from \eqref{H}, 
\be\label{dimlessFriedmann}
\mathcal{H}^2\simeq\alpha^2\left[\left(X-\dot{\Theta}\right)^2+2X/\left(1+\dot{\mathcal{H}}/\mathcal{H}^2\right)\right],
\ee 
which ignores the fermion energy contribution, and defining a parameter which is the dimensionless ratio of the mass scales
\be\label{alpha}
\alpha\equiv\sqrt{\frac{4\pi}{3}}\left(\frac{J}{m^2M_p}\right)\,.
\ee
In the special case ${\bar H}=M$ discussed above after \eqref{g1}, $\,\alpha\simeq 10^3$. The gauge condition \eqref{gauge} reduces to
\be\label{gauge2}
{X}'+3{\mathcal H}X=\Theta\,.
\ee
Taking the derivative of  \eqref{gauge2} and plugging into \eqref{eoms} gives
\be
{\Theta}'=-X+1/\left(1+{\mathcal H}'/{\mathcal H}^2\right)\,.
\ee
Substituting this into \eqref{dimlessFriedmann}, we are able to remove explicit scalar field dependence from the Friedmann equation:
\be\label{H2}
{\mathcal H}^2=\alpha^2\left[3X^2+\left\{X-1/\left(1+{\mathcal H}'/{\mathcal H}^2\right)\right\}^2\right]\, ,
\ee
where ${\mathcal H}>0$ is  taken to be the positive square root, consistent with the desired expansion of the universe.
The dynamics of the system are completely determined by \eqref{H2} and \eqref{eoms} and the parameter
$\alpha$, along with \eqref{gauge2} for the St\"uckelberg field in terms of ${\mathcal H}$ and $X$. For investigating the phase space evolution of
this system, the evolution can be cast in terms of the three
variables $X$, $Y$, and $\epsilon$, the dynamics of which form a system of three first-order differential equations:
\ba\label{sys}
&&{X}'\equiv Y,\nonumber\\
&&{Y}'=(1-\epsilon)^{-1}-X-3{\mathcal H}Y + 3{\mathcal H}^2\epsilon X, \nonumber\\
&&{\epsilon}'=(1-\epsilon)^2\left[Y+\frac{3XY+\alpha^{-2}\epsilon {\mathcal H}^3}{X-\left(1-\epsilon\right)^{-1}}\right],
\ea
where ${\mathcal H}$ on the right sides can be expressed algebraically in terms of $X$ and $\epsilon$ as
\be
\frac{{\mathcal H}^2}{\alpha^2} = 3X^2 + \left(X - \frac{1}{1-\epsilon}\right)^2.
\ee
Here $\epsilon$ is the deceleration parameter defined in \eqref{dec}, and physically $\epsilon=0$ corresponds to a pure de Sitter phase.

\section{Existence and Stability of a Dynamical Attractor}

The fixed points of the dynamical system are found by setting the time derivative of each kinematic variable to zero. The system has exactly one inflationary 
fixed point in the phase space, 
\be\label{fixed}
({\bar X}, {\bar Y}, {\bar \epsilon})=(1,0,0),
\ee
with the fixed-point Hubble parameter ${\bar {\mathcal H}}=\sqrt{3}\alpha$. 
For small departures from the fixed point, define the perturbation variables  $X=1+\delta X$, $Y=\delta Y={\delta X}'$ and $\epsilon=\delta\epsilon$. Upon substituting  this expansion into \eqref{sys}, the second term in the expression for ${\epsilon}'$ has the form
\be
\frac{3XY+\alpha\epsilon\left[3X^2+\{X-\left(1-\epsilon\right)^{-1}\}^2\right]^{3/2}}{X-\left(1-\epsilon\right)^{-1}}\longrightarrow \frac{3\,\delta Y+3\sqrt{3}\,\alpha\epsilon+{\mathcal O}(2)}{\delta X-\delta\epsilon+{\mathcal O}(2)}\,,
\ee
where $\mathcal{O}(2)$ are quantities of second or higher order in the perturbations. This implies that the Jacobian matrix $\mathcal{M}$, defined as 
$\delta\textbf{X}'=\mathcal{M}\,\delta\textbf{X}\,,$
is \emph{not} guaranteed to be finite at the fixed point. Therefore stability {\it cannot} be established via  the standard Hartman-Grobman linearization theorem. 

However, the system is straightforward to evolve numerically. Figure 1 displays several representative trajectories for different initial conditions, setting $\alpha=1$. 
The initial condition must satisfy $\epsilon < 1$, since the current \eqref{Current} is undefined at $\epsilon=1$. Numerous initial conditions have trajectories which
flow to the inflationary fixed point. Also displayed is a trajectory which evolves away from the fixed point and flows to $\epsilon \rightarrow -\infty$. Some results on slices through a grid of initial conditions in three dimensions are displayed in Figure 2. The blue points
represent initial conditions that have been numerically verified to  reach the fixed point \eqref{fixed}. For $X_0>1$, all trajectories are repelled. A
general ``narrowing'' of the attractor basin occurs as $X_0\rightarrow 1$ from below. This behavior looks similar to the convergence of the trajectories as they approach the fixed point in Figure 1. These calculations clearly indicate a substantial attractor basin in the region $X_0<1$. We have verified that the basin remains similarly large for 
$\alpha=10$ and $\alpha = 100$.

\begin{figure}[H]
{\caption{(color online) Phase space trajectories, with initial conditions $(X_0, Y_0, \epsilon_0)$ given in the legend, and for $\alpha=1$. Four trajectories (in colors) starting from $X_0=0$ are attracted to the fixed point, $({\bar X}, {\bar Y}, {\bar \epsilon})=(1, 0, 0)\,$. One trajectory (in gray) starting from $X_0=1.05$ is repelled by the fixed point and has $\epsilon \rightarrow -\infty$.}
\includegraphics[width=6in,bb=0 0 370 280]{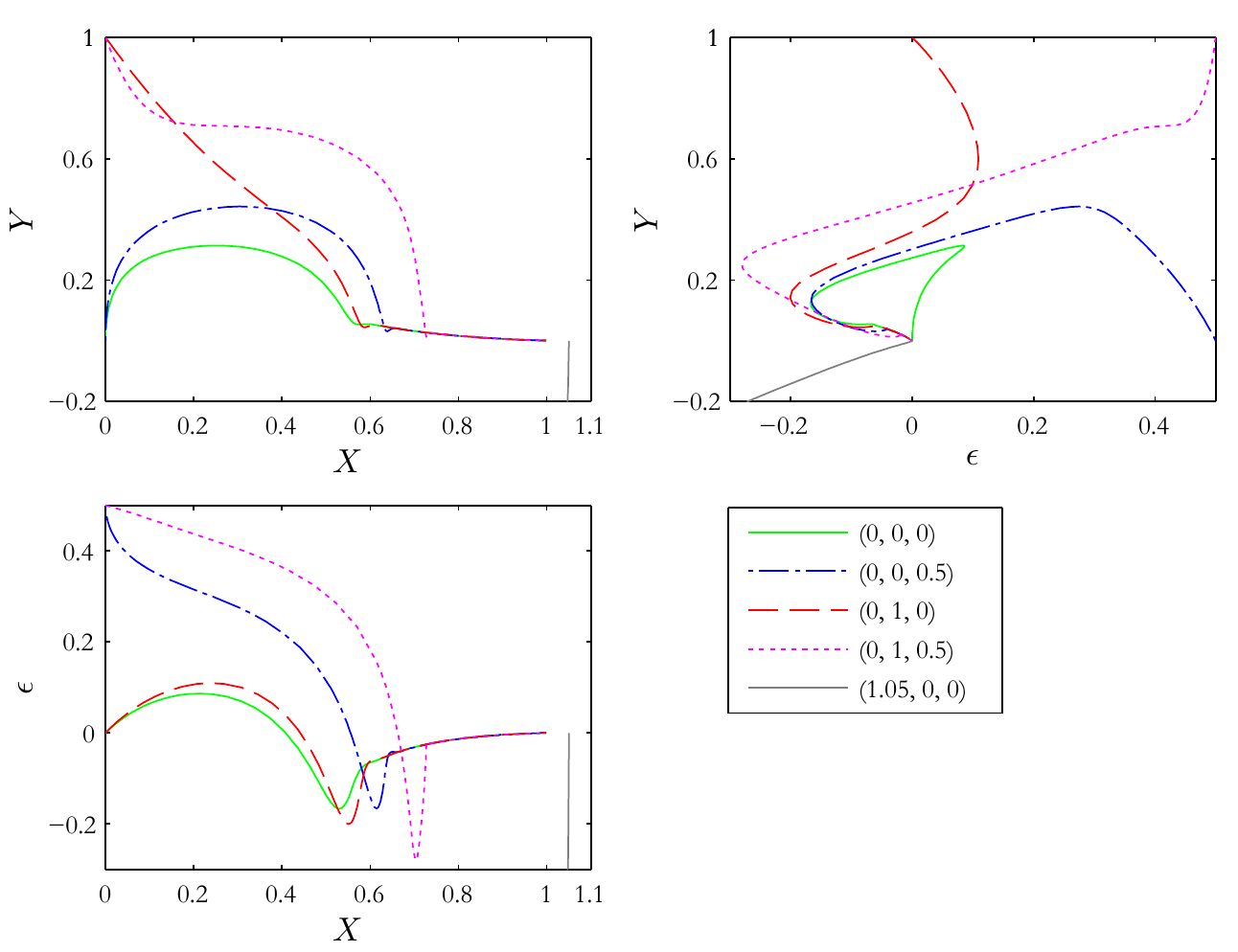}
\label{Figure1}}
\end{figure}

\begin{figure}[H]{\caption {(color online) The stability landscape. Each of these two panels is a 2-dimensional slice of the 3-dimensional phase space. The grid points in blue represent initial conditions which end up at the inflationary fixed point. }
\includegraphics[width=6in,bb=0 0 940 400]{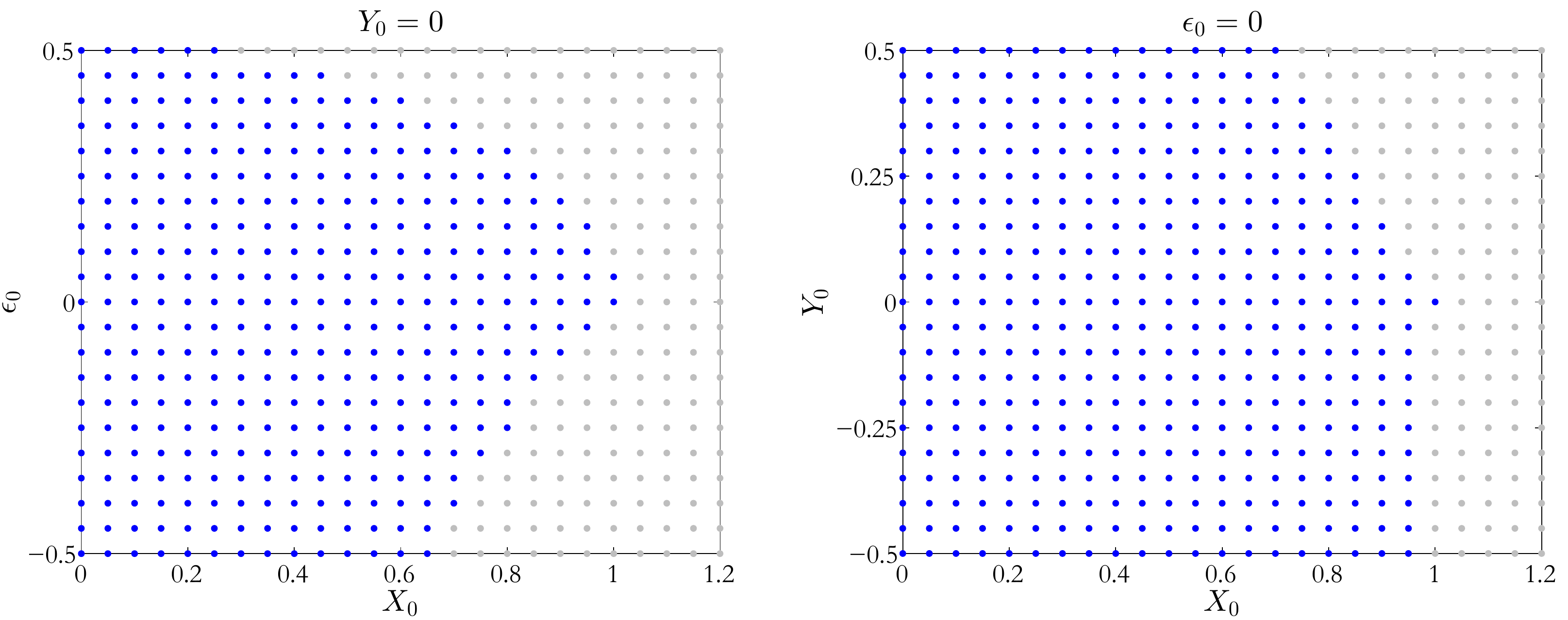}
\label{Figure2}}
\end{figure}

\section{Conclusion and Discussion}
We have demonstrated the existence of self-consistent solutions with inflationary dynamics for a system composed of a fermion current $\mathcal{J}$ coupled to an abelian gauge field $A$ with the standard coupling $A\cdot\mathcal{J}$, with a St\"uckelberg field $\theta$ to guarantee consistency. A similar model in Ref.~\cite{CSI} instead employed a Chern-Simons term. This model differs from the usual scalar field inflation models in that it requires no arbitrary potentials, and uses only standard couplings between fields. 

In this paper, we have shown that the unique dynamical inflation solution is not fine-tuned, in the sense that it arises from a range of homogeneous initial conditions. Inflation is driven by the coupling between the gauge field and the fermion field; the coupling energy density remains constant as the universe expands. The energy density from the fermion current itself is small; the function of fermion current is to drive up the gauge field, which then provides a large energy density via its interaction with the fermion current. This mechanism requires that the fermion field have a bare mass, {\it i.e.} that the vacuum expectation value of $\mathcal{J}^0$ be non-zero. We have demonstrated that for a choice of  fermion mass $M=10^{-3}\, M_p$ and gauge boson mass $m=10^{-6}\, M_p$, we obtain the standard GUT-scale inflation with expansion parameter $H=10^{-3}M_p$. Other combinations of values for the two masses likely also give rise to realistic inflationary solutions, which we leave for future study.

Since we do not postulate a scalar field which by construction has a constant potential energy density as the universe expands, the conceptual issues underlying a particle physics mechanism driving the inflationary behavior come into relief. Fermions must be created as the universe expands in order to maintain a constant fermion current. We have evaluated the vacuum expectation value of the fermion current to demonstrate that such fermion creation occurs in the inflating universe, but we have not explicated a microphysical mechanism for this process. We are currently investigating whether quantum gravitational anomalies which lead to particle creation in space-times with significant curvature \cite{ultimo} can provide sufficient fermion creation rates. The usual problem of initial conditions leading to successful inflation is then replaced by the problem of how the universe came to be in a state with rapid fermion production. While particle production is not conventionally considered in traditional scalar field models of inflation, if the constant field is interpreted as a condensate of scalar particles, some particle creation mechanism is also required. This fact is usually swept under the rug with the assumption of an effective classical scalar field potential. 

We have also assumed that the fermion and gauge fields are strictly homogeneous. These solutions approach a dynamical fixed point, bringing up the old question of how inflation ends. We do not postulate a potential with a field rolling to the minimum; rather the energy driving inflation arises strictly from the particle interactions in the theory. Three logical possibilities for ending inflation exist. First, solutions might exist which approach the dynamical attractor and then move away; this possibility is feasible since the attractor point is not completely stable. However, the universe would need to loiter in the region of the attractor for a relatively long dynamical time without reaching it, and this behavior does not appear to arise generically. A second possibility is that the amplitude $J$ in \eqref{Current} in fact varies slowly in time due to effects arising from non-constant deceleration  \footnote{See previous footnote.}. This variation may impose a characteristic time scale on the length of inflation.  The third possibility is that departures from homogeneity may eventually induce the end of inflation. This possibility is closely related to the issue of how anisotropies are generated, when considered to be variations in the fermion density. If some physical process connected to the fermion production rate and the gauge field coupling imposes a characteristic time scale on the length of inflation, this could provide a natural explanation for why the microwave background appears to violate statistical isotropy on the largest scales \cite{eriksen04,hansen04}, and why it lacks correlations on the largest angular scales (see {\it e.g.} \cite{hinshaw96,spergel03,copi09}). These features are unexplained in standard inflation models which undergo an arbitrarily long period of inflation, other than as statistical flukes. A specific, finite duration of inflation in scalar-field models requires a fine-tuning of the inflaton initial conditions. Another interesting question, related to  how inflation ends in our model, is the evolution of scalar and tensor perturbations. In particular, it would be interesting to investigate if the two chiralities of  tensor modes evolve differently, as was shown to be the case in a non-Abelian gauge field model of inflation in \cite{Jabbari1,Jabbari2}\,.

The model presented here suggests new physical approaches to fundamental elements of inflation: the underlying origin of the effective potential driving inflationary expansion, creation of particles in an expanding space-time, the particle basis for generating perturbations, mechanisms for setting the duration of inflation. We may have the possibility of understanding all of these issues within the context of known physical interactions.

\section*{Acknowledgements} 
\noindent
DJ is grateful to colleague Sam Cormack at Dartmouth for useful discussions, and undergraduate Luis Martinez for assisting with the numerics. AK is partly supported by NSF grant 1312380 through the Astrophysics Theory Program. AM acknowledges hospitality at Dartmouth College, where the project has started, and support from a start-up grant provided by Fudan University. This work used the NASA Astrophysical Data System for bibliographic information.

\end{document}